\documentclass[a4paper,11pt]{article}
\pdfoutput=1 % if your are submitting a pdflatex (i.e. if you have
\usepackage{jcappub} % for details on the use of the package, please
\usepackage[T1]{fontenc} % if needed
\usepackage[utf8]{inputenc}

\usepackage{esvect}

\usepackage{slashed}

\usepackage{subcaption}

\renewcommand{\O}{\Omega}

\usepackage{xspace}
\newcommand*{\ie}{i.e., }
\newcommand*{\eg}{e.g., }
\newcommand*{\fig}{Fig.\@\xspace}
\newcommand*{\Eq}{Eq.\@\xspace}
\newcommand*{\Eqs}{Eqs.\@\xspace}

\usepackage{xifthen}% Provides \isempty test
\newcommand*\diff{\mathrm{d}} % Straight differential
\newcommand*\ldiff[2][]{ \ifthenelse{\isempty{#1}}{ \diff #2}{\diff^#1#2} \,} % Differential with measure; the mandatory argument is the name of the measure, the option one is the dimension
\let\limitint\int % Only when I provide explicit limits for the integration, I need to do the spacing myself
\renewcommand{\int}{\limitint \!} % The standard integral should have correct spacing

\title{Quantum Effects in Palatini Higgs Inflation}
\author[a]{Mikhail Shaposhnikov,}
\author[a]{Andrey Shkerin,}
\author[a]{Sebastian Zell}
\affiliation[a]{Institute of Physics, Laboratory for Particle Physics and Cosmology, \\
		\'Ecole Polytechnique F\'ed\'erale de Lausanne, CH-1015 Lausanne, Switzerland}
\emailAdd{mikhail.shaposhnikov@epfl.ch}
\emailAdd{andrey.shkerin@epfl.ch}
\emailAdd{sebastian.zell@epfl.ch}

\abstract{
We study quantum effects in Higgs inflation in the Palatini formulation of gravity, in which the metric and connection are treated as independent variables. We exploit the fact that the cutoff, above which perturbation theory breaks down, is higher than the scale of inflation. Unless new physics above the cutoff leads to unnaturally large corrections, we can directly connect low-energy physics and inflation. On the one hand, the lower bound on the top Yukawa coupling due to collider experiments leads to an upper bound on the non-minimal coupling of the Higgs field to gravity: $\xi \lesssim 10^8$. On the other hand, the Higgs potential can only support successful inflation if $\xi \gtrsim 10^6$. This leads to a fairly strict upper bound on the top Yukawa coupling of $0.925$ (defined in the $\overline{\text{MS}}$-scheme at the energy scale $173.2\,\text{GeV}$) and constrains the inflationary prediction for the tensor-to-scalar ratio. Additionally, we compare our findings to metric Higgs inflation.
}

\begin{document}
\maketitle
\flushbottom

\section{Introduction}
\label{sec:introduction}

Among the greatest advances of high-energy physics in recent years are the discovery of the Higgs boson at LHC \cite{1207.7214, 1207.7235}, as it was predicted more than 50 years ago \cite{Englert:1964et, Higgs:1964ia, Higgs:1964pj}, and measurements of the cosmic microwave background (CMB) with unprecedented precision \cite{1001.4538, 1807.06211}. The latter are fully compatible with the paradigm of inflation \cite{Starobinsky:1980te, Guth:1980zm, Linde:1981mu, Mukhanov:1981xt}. It is exciting to think that the Higgs boson and inflation are connected, namely that the former serves as the inflaton. This proposal  \cite{0710.3755} stands out among inflationary models since it does not require any new degrees of freedom beyond the Standard Model.

Higgs inflation is known in  two incarnations --  in the original one suggested in \cite{0710.3755} and its Palatini variant proposed in \cite{0803.2664}.\footnote
{See \cite{1807.02376} and \cite{2001.10135} for reviews of metric and Palatini Higgs inflation, respectively.}
Both theories are described by the same action
\begin{equation} \label{action_full}
S= \int \ldiff[4]{x} \sqrt{-g} \left\{-\frac{M_P^2+2\xi H^\dagger H}{2}R  \right\} +S_{\text{SM}} \,,  
\end{equation}
where $S_{\text{SM}}$ is the Standard Model part, $H$ is the Higgs field and  $M_P=2.435\times10^{18}\,$GeV is the reduced Planck mass. The parameter $\xi\gg1$ sets the strength of the non-minimal coupling between the Higgs field and gravity.\footnote
	{Note that even if there were no such coupling at tree level, it would be generated by quantum corrections \cite{Callan:1970ze}.
	}
  The two versions of Higgs inflation differ in the way the gravitational degrees of freedom are treated. In \cite{0710.3755}, the metric formulation of gravity was used, \ie the dynamical variable is $g_{\mu\nu}$, whereas in the Palatini formulation \cite{Palatini19,Einstein25}, both the metric and connection $\Gamma^\rho_{\mu\nu}$ are treated as independent variables. If $\xi$ is zero, both formulations are equivalent, but this changes once a non-minimal coupling is present. Therefore, the action \eqref{action_full} treated in the Palatini formulation of gravity leads to different physical results, as was shown in \cite{0803.2664}. We shall refer to it as \textit{Palatini Higgs inflation}. In contrast, we call the original model \textit{metric Higgs inflation.}
  
Apart from $\xi$, Higgs inflation is sensitive to another dimensionless parameter, the quartic coupling $\lambda$ of the Higgs field. In the metric theory, it turns out that after taking into account the amplitude of CMB fluctuations, the predictions for the spectral index $n_s$ and the tensor-to-scalar ratio $r$ are solely sensitive to the number of e-foldings $N$.
The Palatini model yields the same functional form of the spectral index, and the resulting numerical value is only slightly different (because $N$ is lower). In contrast, the tensor-to-scalar ratio is suppressed by $\xi$ \cite{0803.2664}.
The predictions of both metric and Palatini inflation are fully compatible with current CMB observations \cite{1807.06211}.\footnote
{See \cite{1303.3787} for a review of various inflationary models and their compatibility with measurements.}

The classical physics picture is modified by quantum effects. First, the coupling constants of the Standard Model, in particular $\lambda$, depend on energy and thus must be evaluated at inflationary scales. This does not only change their numerical values but also the shape of the inflationary potential. Second, the theory is non-renormalizable. Because of the longitudinal gauge bosons of the Standard Model, this is true even if transverse gravitational degrees of freedom -- gravitons -- are disregarded \cite{1008.5157}.

The non-renormalizability results in the appearance of a new energy scale $\Lambda$ in the theory, sometimes called the ultraviolet cutoff. It is derived as follows. Take some particular Higgs field background $h= \sqrt{2 H^\dagger H}$ and consider scattering of different quanta at energy $E$. Since the theory is non-renormalizable, the tree-level cross sections in it increase with energy and hit the unitarity bound at $E=\Lambda$, signaling the breakdown of the perturbative description. In vacuum, \ie for $h=0$, the cutoff is
\begin{equation} \label{cutoff}
\Lambda_{\text{metric}}(h)|_{h\to 0} \simeq \frac{M_P}{\xi}\,, \qquad \qquad  \Lambda_{\text{Palatini}} (h)|_{h\to 0}\simeq \frac{M_P}{\sqrt{\xi}}\,,
\end{equation} 
for the metric case \cite{0902.4465,0903.0355}  and the Palatini one \cite{1012.2900}, respectively.\footnote
{Also in the metric theory of gravity, there are proposals to increase $\Lambda(0)$ without introducing new degrees of freedom \cite{1003.2635, 1005.2978}. They rely on higher-dimensional operators. The perturbative cutoff in the theory studied in \cite{1003.2635} was further discussed in \cite{1612.06253, 1711.08761}.}
In contrast, the inflationary value of the potential has the same dependence on $\xi$ in both theories, $U\sim \lambda M_P^4/\xi^2$, so that $(U)^{\frac{1}{4}}\gg \Lambda_{\text{metric}}(0)$ and $(U)^{\frac{1}{4}}\lesssim \Lambda_{\text{Palatini}}(0)$. Clearly, one can only trust the predictions of the classical picture outlined above if all relevant energy scales $E_{\rm inf}\approx (U)^{\frac{1}{4}}$ are below $\Lambda(h)$.  

For investigation of the self-consistency of the tree analysis {\em during inflation}, taking place at $h \neq 0$, one needs to repeat the cutoff computation for the non-zero background of the Higgs field for the action (\ref{action_full}). This reveals the field dependence \cite{1008.5157} of the cutoff (see also \cite{Ferrara:2010in}). It so happens that during inflation, the condition  $E_{\rm inf}< \Lambda(h)$ is fulfilled for both the metric and the Palatini scenarios. This relies on the fact that in the metric theory (\ref{action_full}) the cutoff scale increases as $h$ becomes larger \cite{1008.5157}.\footnote{It is still an open question whether there exists a UV-completion of the theory (\ref{action_full}) that realizes the field-dependent cutoff found in \cite{1008.5157} (see also below).}

The analysis of preheating in metric Higgs inflation indicates that the temperature of the Universe {\em after inflation} can be larger than the cutoff \cite{1609.05209, 1610.08916}. This does not spoil the predictions for $n_s$ and $r$ \cite{1008.5157, 1412.3811}, but tells us that there is a period in the history of the Universe which is not treated in a consistent way. Straightforward extrapolation indicates that preheating happens instantaneously because of parametric resonance creating longitudinal vector bosons \cite{1609.05209, 1610.08916}.\footnote
{For earlier studies of preheating in metric Higgs inflation see \cite{0812.3622, 0812.4624}.}
There is no such problem in the Palatini scenario, since the associated temperature lies below the cutoff. Preheating turns out to be instantaneous due to the tachyonic production of Higgs excitations \cite{1902.10148}.

The question about what happens above the cutoff scale $\Lambda(h)$ remains. One possibility, on which we shall not focus, is that new degrees of freedom appear. This was explored in \cite{1010.1417, 1307.5298, 1501.02231, 1701.07665} for the metric scenario. In those models, the theory stays in the weak coupling regime at energies above $\Lambda(h)$.\footnote
{In the models \cite{1010.1417, 1307.5298, 1501.02231, 1701.07665}, the perturbative cutoff is field-independent. Thus, the scale at which the new degrees of freedom decouple is given by $\Lambda(0)$ for all values of $h$.}
The price to pay is that a degree of freedom beyond the Higgs boson participates in inflationary dynamics, i.e., the Higgs inflation is modified at tree level.  Another possibility is the idea of ``self-healing'' \cite{1008.5157,Aydemir:2012nz}: there are no new fundamental degrees of freedom, and the theory just enters the regime of strong coupling between the known constituents of the Standard Model. Moreover, one assumes a minimal UV-completion which amounts to adding to the Lagrangian only those operators that are needed to remove divergencies. Hence, the inflationary dynamics remains unchanged at tree level. As finite parts of the resulting counter-terms cannot be fixed by the requirement of the finiteness of the amplitudes, they stay arbitrary. Eventually, they will be fixed by the construction of a UV complete theory, in analogy with the low-energy chiral theory describing pions, being an effective theory of QCD. This hypothesis was proposed in \cite{1008.5157} and worked out in more detail in \cite{1412.3811}. It provides a minimal and well-defined setup for dealing with non-renormalizable theories without adding new degrees of freedom. 

It is thus far unknown how the two scenarios of Higgs inflation will ultimately be UV-completed. Therefore, one cannot tell which of the two possibilities --  weak coupling with new degrees of freedom or strong coupling without new particles -- is realized. The plausibility of both options has been extensively discussed in the literature, see \cite{0902.4465, 0903.0355,Ferrara:2010in, 1008.5157, 1010.1417, 1012.2900, 1501.02231}. In the following, we will explore the second option, \ie we assume a minimal UV-completion that does not introduce new degrees of freedom and leaves Higgs inflation unaltered at tree level.

In this scenario, it is interesting to investigate if the inflationary value of $\lambda$ can be derived from the low-energy parameters of the Standard Model as measured at collider experiments. The relevant energy scale for the evaluation of the corresponding RG evolution is on the order of the top quark mass during inflation, $\mu_{\text{inf}}\sim y_t M_P/\sqrt{\xi}$, where $y_t$ is the high-energy value of the top Yukawa coupling. For metric Higgs inflation, this question has been studied in detail for the ``self-healing'' setup \cite{1008.5157,1403.6078,1412.3811,Fumagalli:2016lls,Enckell:2016xse, 1706.05007}. It was shown that the effect of adding necessary counter-terms results in ``jumps'' of the different coupling constants. In addition, new coupling parameters, which were absent at low energies, appear at larger field values \cite{0904.1537}. Thus, the RG evolution of $\lambda$ as computed within the Standard Model is modified by an a priori arbitrary constant $\delta \lambda$. In other words, the relation between the low- and high-energy parameters is in general lost. This matches the fact that $\Lambda(0) < \mu_{\text{inf}}$ in the metric scenario. Nevertheless, one can explore the consequences of assuming that $\delta \lambda$ vanishes. In this case, the inflationary potential is determined by the low-energy parameters of the Standard Model. Eventually, this leads to an upper bound on the low-energy value $y_t^{\text{low}}$ of the top Yukawa coupling beyond which Higgs inflation becomes impossible \cite{0812.4950, 0904.1537,1411.1923}.\footnote
{See also \cite{0809.2104, 0812.4946, 0904.1698, 0910.1041} for early studies of quantum effects in Higgs inflation.}

One of the aims of the present work is to extend this analysis to Palatini Higgs inflation, i.e., we assume that the RG evolution of the Standard Model is valid up to inflationary scales and then derive a bound on the low-energy value of the top Yukawa coupling from the requirement that Palatini Higgs inflation can successfully take place.\footnote
{Other aspects of quantum effects in Palatini Higgs inflation have been considered in \cite{1709.07853,1712.04874, 1802.09299}.}
We will see that the main conclusions of the metric case remain in force, but the upper bound on $y_t^{\text{low}}$ is somewhat relaxed. Within the same assumptions and with the use of experimental constraints on $y_t^{\text{low}}$, we also determine the range of values of the non-minimal coupling $\xi$, thus constraining the inflationary parameter $r$ in the Palatini scenario.

An advantage of Palatini Higgs inflation is that its cutoff $\Lambda(0)$, above which perturbation theory breaks down, is higher than inflationary energy scales \cite{1012.2900}. In particular, it lies slightly above the energy $\mu_{\text{inf}}$, which is relevant for the evaluation of the running couplings. This raises the hope that the assumption of validity of the Standard Model running up to inflationary scales  is less strong, \ie can be justified more easily, in the Palatini scenario. We will attempt to quantify this statement. Within the scenario of a minimal UV-completion, we conclude that, unless corrections due to unknown UV physics above $\Lambda(0)$ are unnaturally large, the connection of low- and high-energy parameters can indeed be established.
In summary, the bound on $y_t^{\text{low}}$ turns out to be similar in metric and Palatini Higgs inflation, but it is more robust against corrections due to new physics in the latter theory.

This paper is organized as follows. In section \ref{sec:review}, we provide a short overview of metric and  Palatini Higgs inflations at the classical level. In section \ref{sec:newPhysics}, we discuss the modifications of the classical picture due to quantum effects and estimate the influence of new physics above the cutoff scale. We will see that, depending on the type of new physics and the value of $y_t^{\text{low}}$, the relationship between low- and high-energy parameters may be modified or stay intact. Assuming that corrections to the Standard Model running of $\lambda$ are small, we establish the connection of high- and low-energy physics  and obtain the upper bound on the top quark Yukawa coupling in section \ref{sec:connection}. We discuss our results in section \ref{sec:conclusion}.

\section{Metric vs Palatini: classical picture}
\label{sec:review}

We review Higgs inflation both in its metric \cite{0710.3755} and Palatini version \cite{0803.2664}. We focus on the Higgs-gravity sector of the action (\ref{action_full}):
\begin{equation}\label{action}
S_{\text{grav.+h.}}=\int \ldiff[4]{x} \sqrt{-g} \left\{ -\frac{M_P^2+\xi h^2}{2} R + \frac{1}{2} (\partial_\mu h)^2 -\frac{\lambda}{4}h^4\right\}\,,
\end{equation}
where we chose the unitary gauge for the Higgs field, $H=(0,h)^T/\sqrt{2}$. Here and below we neglect the Higgs field vacuum expectation value $v$. This is justified when considering the inflationary epoch with $h\gg v$. To compare the metric and Palatini scenario, it is convenient to write the theory in a form without the non-minimal coupling of $h$ to the Ricci scalar. 
To this end, we perform a conformal transformation of the metric:
\begin{equation}\label{FieldTransform1}
\hat{g}_{\mu\nu}=\Omega^2g_{\mu\nu} \,,~~~ \Omega^2=1+\frac{\xi h^2}{M_P^2} \,.
\end{equation}
The action (\ref{action}) becomes
\begin{equation} \label{einsteinNonCanonical}
S_{\text{grav.+h.}}= \int \ldiff[4]{x} \sqrt{-\hat{g}} \left\{-\frac{M_P^2}{2} \hat{R} + \frac{1}{2}K(h) (\partial_\mu h)^2 -\frac{\lambda}{4\Omega^4}h^4\right\}\,.
\end{equation}
Since we have removed the non-minimal coupling, the two formulations of gravity are equivalent from here on. However, the function $K(h)$ is different in the two theories, owing to the fact that in the Palatini case the Ricci tensor $R_{\mu\nu}$ does not depend on the metric and hence is not transformed under (\ref{FieldTransform1}). The explicit form of $K(h)$ is presented in table \ref{table}. Next, we make the kinetic term of the scalar field canonical by introducing the field $\chi$ via
\begin{equation}\label{FieldTransform2}
\frac{dh}{d\chi}=\frac{1}{\sqrt{K(h)}} \,.
\end{equation}
In the Palatini case, there is a simple closed expression for $h(\chi)$ whereas in the metric scenario we rely on an approximation that is valid for large field values (see table \ref{table}).\footnote
	{Also in the metric scenario, \Eq \eqref{FieldTransform2} can be integrated exactly, giving $\chi$ as a function of $h$ \cite{0812.4624}.}
Rewriting the Lagrangian in terms of $\hat{g}_{\mu\nu}$ and $\chi$, we arrive at
\begin{equation} \label{einsteinAction}
S_{\text{grav.+h.}} = \int \ldiff[4]{x} \sqrt{-\hat{g}} \left\{-\frac{M_P^2}{2} \hat{R} + \frac{1}{2}(\partial_\mu \chi)^2 - U(\chi) \right\}\,,
\end{equation}
where
\begin{equation}\label{Potential}
U(\chi)=\dfrac{\lambda}{4}F(\chi)^4 \,,
\end{equation}
and the function $F(\chi)$ is again different in the two theories, see table \ref{table}. At large field values ($\chi>M_P$ and $\chi > M_P/\sqrt{\xi}$ for the metric and Palatini scenario, respectively), the potential $U(\chi)$ flattens and allows for inflation. Whereas the potential at inflationary energies scales as $U(\chi)\sim M_P^4\lambda/\xi^2$ in both models, inflation happens at a lower scale in the Palatini case because $\xi$ turns out to be larger than in the metric theory.

	\begin{table}
	\begin{tabular}{c|c@{\hskip 0.1\linewidth}c}
		& Metric 							 & Palatini 		\\
		\hline \\
		$K(h)$		 & $\dfrac{1}{\O^2}+\dfrac{6\xi^2h^2}{M_P^2\O^4}$ & 	$\dfrac{1}{\O^2}$ 		\\
		\\
		$h(\chi)$ & $\frac{M_P}{\sqrt{\xi}} \exp \frac{\chi}{\sqrt{6}M_P}$ ~~ for ~~  $h\gg \frac{M_P}{\sqrt{\xi}}$ & $\dfrac{M_P}{\sqrt{\xi}}\sinh\dfrac{\sqrt{\xi}\chi}{M_P}$\\  \\
		$F(\chi)$	 & $\begin{cases}
		\chi & \chi<\frac{M_P}{\xi}\\
		\sqrt{\frac{\sqrt{2}M_P \chi}{\sqrt{3}\xi}}& \frac{M_P}{\xi}<\chi <M_P\\
		\frac{M_P}{\sqrt{\xi}}\left(1-e^{-\sqrt{2/3}\chi/M_P}\right)^{1/2} & M_P< \chi  \end{cases}$ & $\dfrac{M_P}{\sqrt{\xi}}\tanh\dfrac{\sqrt{\xi}\chi}{M_P}$ \\ \\
			$T_{\text{reh}} $		 &  $ 3 \cdot 10^{15} \, \text{GeV}$				 & $\approx 4 \cdot 10^{13} \, \text{GeV}$\\ \\
			$N$	 & $55.4$ & $54.9-\dfrac{1}{4}\ln\xi \approx 50.9$\\ \\
		$n_s$		 & $1-\dfrac{2}{N} = 0.964$ 				 & $1-\dfrac{2}{N} \approx 0.961$ \\ \\
		$r$		 & $\dfrac{12}{N^2} = 3.9\cdot 10^{-3}$			 & $\dfrac{2}{\xi N^2}\approx 7.7\cdot 10^{-11}$ 
	\end{tabular}
	\caption{Comparison between metric and Palatini Higgs inflation at tree level and for $\xi\gg 1$. In the Palatini case, we use $\xi = 10^7$. Since the analysis of this paper shows that $\xi$ can deviate from this value by an order of magnitude, we use the symbol ``$\approx$'' when displaying numerical values that depend on $\xi$. Expressions for $n_s$ and $r$ are given to the leading order in $N^{-1}$ and $\xi^{-1}$. }
	\label{table}
\end{table}

As a side remark, it is possible to obtain the action (\ref{einsteinAction}) of Palatini Higgs inflation without using the Palatini formulation of gravity. To this end, it suffices to add the term 
\begin{equation}
\delta S = -\int \ldiff[4]{x} \sqrt{-g}\frac{3\xi^2}{M_P^2 \Omega^2}h^2(\partial_\mu h)^2
\end{equation}
to the action (\ref{action}) in the metric formulation. This term cancels the contribution due to the conformal transformation of $R_{\mu\nu}$. Thus, although the metric and Palatini formulations of gravity lead to different predictions for Higgs inflation, one cannot use this difference to test the nature of gravitational degrees of freedom. Instead, one only determines in which variables the theory looks simpler.

Let us now compare inflationary observables in the two models. We introduce the slow-roll parameters
\begin{align} \label{slowRollParameters}
\epsilon =  \frac{M_P^2}{2} \left(\frac{\diff U/\diff \chi}{U}\right)^2  \,, \qquad \eta = M_P^2 \frac{\diff^2 U/\diff^2 \chi}{U} \,,
\end{align}
and it is straightforward to compute them as functions of the field $h$. In turn, we can express the latter in terms of the number of e-foldings $N$:
\begin{equation} \label{NofH}
	N = \frac{1}{M_P^2}\limitint_{h_{\text{end}}}^{h} \frac{U}{\diff U/\diff \chi} \frac{\diff \chi}{\diff h} \diff h \,,
\end{equation}
where $h_{\text{end}}$ is the field value at the end of inflation.  It is determined by the condition that either $\epsilon$ or $|\eta|$ becomes of the order of $1$. Next we compute the number of e-foldings, at which CMB perturbations are generated. As input, we use the Planck normalization \cite{1807.06211},
\begin{equation} \label{normalization}
	\dfrac{U}{\epsilon} = 5.0\cdot 10^{-7} M_P^4 \,,
\end{equation}
at the scale $k_\star=0.05\,$Mpc$^{-1}$.  Moreover, we approximate preheating as instantaneous in both models \cite{1609.05209, 1610.08916, 1902.10148}, as explained in the introduction. Then, the preheating temperature is given by
\begin{equation}
	T_{\text{reh}}= \left(\frac{30 U}{\pi^2 g_\star}\right)^{1/4} \,,
\end{equation}
where $g_\star=106.75$ is the effective number of degrees of freedom at the moment of preheating.  Since $U$ is smaller in the Palatini case than in the metric case, $T_{\text{reh}}$ is correspondingly lower.\footnote
	{Because of its smaller preheating temperature, Palatini Higgs inflation is favorable for QCD axion models in which the Peccei-Quinn symmetry is broken before the end of inflation \cite{1906.11837}.}
In turn, the preheating temperature allows us to determine $N$ \cite{1902.10148}. The resulting values for $T_{\text{reh}}$ and $N$ are shown in table \ref{table}.

We see that the leading-order predictions of metric Higgs inflation are independent of both $\xi$ and $\lambda$, \ie they are fully fixed by the CMB normalization \eqref{normalization}. In contrast, there is one free parameter ($\xi$ or equivalently $\lambda$) in the Palatini scenario. It cannot be determined at the classical level, in particular, without knowing how $\lambda$ runs. Our subsequent analysis will show that $\xi$ lies between $\sim 10^6$ and $\sim 10^8$. Therefore, we use $\xi=10^7$ as a typical value. Changing $\xi$ by an order of magnitude alters $N$ by about $0.5$.
Finally, we compute the spectral tilt, $n_s=1-6\epsilon+2\eta$, as well as the scalar-to-tensor ratio, $r=16\epsilon$, which are also displayed in table \ref{table}. We see that in the metric theory both indices only depend on $N$, while in the Palatini case $r$ is suppressed by one power of $\xi$. Both in the metric and Palatini scenarios, the resulting values are fully compatible with recent CMB measurements \cite{1807.06211}. 

\section{Metric vs Palatini: quantum corrections}
\label{sec:newPhysics}

We have reviewed the two versions of Higgs inflation at tree level. A consistent analysis of inflation is, however, impossible without considering quantum corrections, since they modify the inflationary potential via the running of the Higgs self-coupling $\lambda$. 
This leads to the question if the corresponding RG evolution can be computed within the Standard Model. The relevant energy scale $\mu_{\text{inf}}$ for evaluating it is on the order of the top quark mass during inflation:\footnote
{This well-known choice follows from the principle of minimal sensitivity \cite{Stevenson:1981vj,Stevenson:1982wn}, because the top quark has the strongest influence on the running of $\lambda$. Nevertheless, normalizing to the Higgs mass during inflation, $\mu=m_h\approx \sqrt{\lambda} F(\chi)$, could be more appropriate in certain computations. Since $m_h<y_t F(\chi)$, this choice would only decrease our subsequent estimates of the influence of new physics above $\Lambda$.}
\begin{equation} \label{mu}
	\mu_{\text{inf}} = \mu(\chi)= y_t F(\chi) \,.
\end{equation}
As is evident from table \ref{table}, both the metric and the Palatini theories yield $\mu_{\text{inf}}\approx y_t M_P/\sqrt{\xi}$ at inflationary energies. For the high-energy value of the top Yukawa coupling, we use\footnote
{We obtain this result by evaluating the value of $y_t$ at the energy scale $M_P/\sqrt{10^7}$, starting from its low-energy value $y_t^{\text{low}} = 0.92354$; see the discussion below \Eq \eqref{runningCoupling}.}
\begin{equation} \label{Yt}
	y_t = 0.43 \,,
\end{equation}
and we neglect the further running of $y_t$ at inflationary scales. We need to compare $\mu_{\text{inf}}$ with the perturbative cutoff $\Lambda$. For the evaluation of the RG equation, the cutoff in vacuum is relevant. Therefore, $\Lambda$ refers to $\Lambda(0)$ in the following. In metric Higgs inflation, we have $\mu_{\text{inf}}\gg \Lambda$, whereas the Palatini scenario leads to $\mu_{\text{inf}}\lesssim \Lambda$ (see \Eq \eqref{cutoff}). Thus, we expect that Palatini inflation is more robust against corrections due to new physics above $\Lambda$. Our goal is to quantify this statement.

As explained before, we work under the assumption that UV-physics does not modify the inflationary predictions at tree level. In principle, this could be realized in two different setups. In the first one, new degrees of freedom appear around $\Lambda$. In the second option, the breakdown of perturbation theory at the cutoff is merely an indication of strong coupling among the known constituents of the Standard Model \cite{1008.5157}. Since all explicit proposals of a UV-completion known to us \cite{1010.1417, 1307.5298, 1501.02231, 1701.07665} that rely on introducing new particles alter inflation at tree level, the second option appears to be more favorable. In this case, the assumption of a minimal UV-completion amounts to excluding higher-dimensional operators that do not respect the approximate shift symmetry $\chi \rightarrow \chi + $const of the potential $U(\chi)$.\footnote
{Adding those can easily make inflation impossible (or change inflationary predictions) \cite{1712.08471, 1904.05699}. In contrast, quantum corrections that are generated by the theory itself are small \cite{1712.04874}.}

First, we employ the approach of \cite{1008.5157,1412.3811}. It relies on adding to the theory only those operators that are needed to remove divergencies. Beyond that, it does require any assumptions about the behavior of the theory at $\Lambda$. In \cite{1008.5157,1412.3811}, only metric Higgs inflation was studied, but we can analogously apply the analysis to the Palatini case, keeping in mind that the function $F(\chi)$ is different. The starting point is the computation of one-loop contributions to the effective Higgs potential, in the background field $\chi$ and in dimensional regularization \cite{tHooft:1972tcz}.
Following \cite{1412.3811}, the top quark contribution gives 
\begin{align}
\begin{minipage}[h]{0.1\linewidth}
\center{\includegraphics[width=0.7\linewidth]{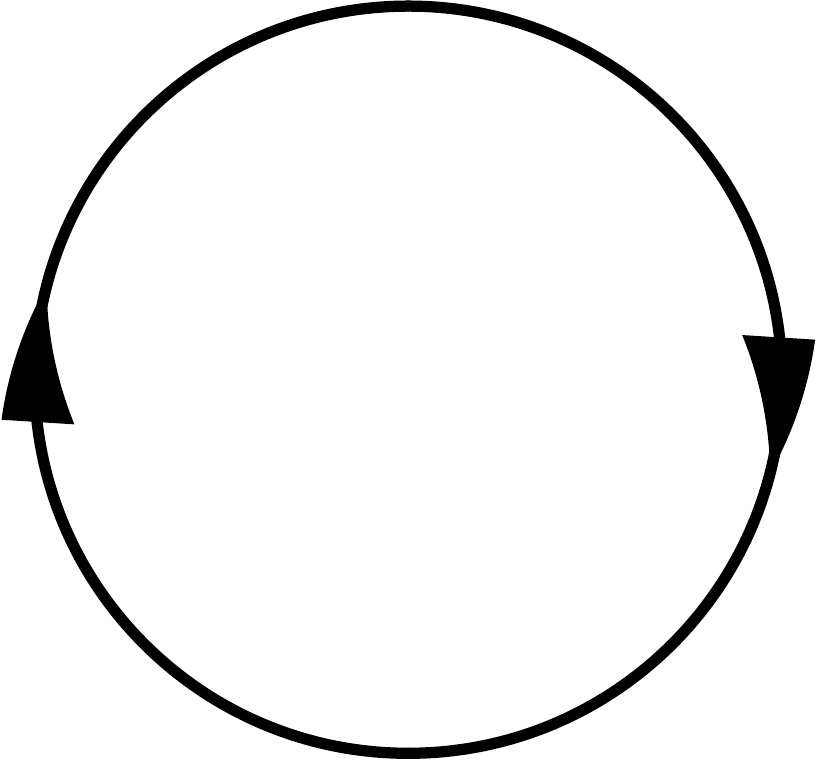}}
\end{minipage}
& =-  \text{Tr} \ln \left(i \slashed{\partial} + y_t F\right) \nonumber\\
& = -\frac{y_t^4}{64 \pi^2} \left(\frac{2}{\bar{\epsilon}} -\ln \frac{y_t^2 F^2}{2 \mu^2} +\frac{3}{2}\right) F^4 \,,
\end{align}
and the Higgs loop is
\begin{align}
\begin{minipage}[h]{0.1\linewidth}
\center{\includegraphics[width=0.7\linewidth]{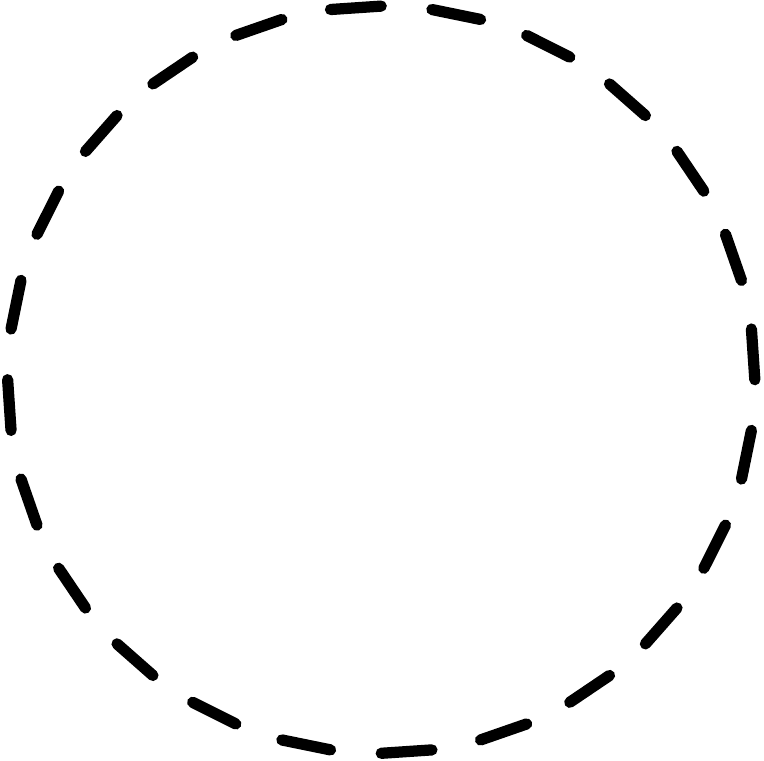}}
\end{minipage}
& = \frac{1}{2} \text{Tr} \ln \left[\Box - \left(\frac{\lambda}{4}(F^4)''\right)^2\right] \nonumber\\
& = \frac{9 \lambda^2}{64 \pi^2} \left(\frac{2}{\bar{\epsilon}} -\ln \frac{\lambda (F^4)''}{4 \mu^2} +\frac{3}{2}\right)\left(F^{'2} + \frac{1}{3} F''F\right)^2 F^4 \,.
\end{align}
We denoted $2/\bar{\epsilon} = 2/\epsilon-\gamma + \ln 4\pi$, and the fractional dimension is $D=4-2\epsilon$.

To cancel the divergences in the loops, counter-terms must be introduced \cite{1412.3811}:
\begin{equation} \label{counterTerms}
\delta \mathcal{L}_{\text{ct}} = \left(-\frac{2}{\bar{\epsilon}}\frac{9 \lambda^2}{64 \pi^2} + \delta \lambda\right) \left(F^{'2} + \frac{1}{3} F''F\right)^2 F^4 + \left(\frac{2}{\bar{\epsilon}}\frac{y_t^4}{64 \pi^2} - \delta\tilde{ \lambda}\right)F^4 \,.
\end{equation}
The divergent terms in the Lagrangian are uniquely determined by the requirement to cancel the corresponding contributions from the loops. In contrast, the finite contributions, which are proportional to $\delta {\lambda}$ and $\delta\tilde{\lambda}$, cannot be determined by the present argument.
We see that the constant $\delta\tilde{\lambda}$ can be eliminated by a redefinition of $\lambda$, but this is not possible for $\delta\lambda$. The reason is that for low values of $\chi$ the quantity
\begin{equation}
\mathcal{F}=F^{'2} + \dfrac{1}{3} F''F
\end{equation}
is on the order of one, and $\delta \lambda$ contributes to the effective value of $\lambda$. In contrast, for large values of $\chi$ the coefficient $\mathcal{F}$ vanishes, and the effect of $\delta \lambda$ disappears. This is true both in the metric and in the Palatini scenario. One can parametrize this behavior by replacing
\begin{equation} \label{lambdaReplacement}
\lambda(\mu) \rightarrow 	\lambda(\mu) + \left(\mathcal{F}^2-1\right) \delta \lambda\,,
\end{equation}
where $\lambda(\mu)$ is evaluated by using the Standard Model diagrams only. If $\delta \lambda$ is comparable to or larger than $\lambda(\mu)$, this means that the inflationary value of $\lambda$ cannot be determined from the RG evolution as computed within the Standard Model. In contrast, if $\delta \lambda$ is much smaller, \eg on the order of magnitude of the coefficient of the divergent term in \Eq \eqref{counterTerms}, the connection of high- and low-energy physics can be established. Again, this is true both in metric and Palatini Higgs inflation. Thus far, however, we have not made a statement about the order of magnitude of $\delta \lambda$ in the two theories.

As a next step, we study at which field value the function $\mathcal{F}$ deviates considerably from $1$. This would correspond to the point at which a ``jump'' in the coupling $\lambda$ occurs. Using the expressions for $F(\chi)$ presented in table \ref{table}, we get $\chi_{\text{jump}}\sim M_P/\xi$ in the metric case and $\chi_{\text{jump}}\sim M_P/\sqrt{\xi}$ in the Palatini case. We need to compare these values to the magnitude of $\chi$ during inflation. Using \Eqs (\ref{FieldTransform2}) and (\ref{NofH}), we get $\chi_{\text{inf}}\sim M_P \ln N$ and $\chi_{\text{inf}}\sim M_P / \sqrt{\xi} \ln (\xi N)$, respectively. Thus, we have $\chi_{\text{inf}}> \chi_{\text{jump}}$ in both scenarios, so that inflationary parameters can in principle be sensitive to new physics in both theories. However, the values of $\chi_{\text{inf}}$ and $\chi_{\text{jump}}$ are much closer in the Palatini case. For this reason, the latter theory is less sensitive to corrections due to new physics above $\Lambda$, as we will show more explicitly shortly.
	
Before that, we remark that we can arrive at the same conclusion by arguing in terms of the energy scale $\mu_{\text{jump}}$, at which the jumps occurs. As is evident from \Eq \eqref{mu} and the functions $F(\chi)$ presented in table \ref{table}, we have $\mu_{\text{jump}} \sim y_t \chi_{\text{jump}}$ in both theories. Since the functional form of the inflationary value $\mu_{\text{inf}}$ does not distinguish between the two scenarios, we again find that $\mu_{\text{jump}}$ lies parametrically below $\mu_{\text{inf}}$ in the metric case, while the two quantities are of the same order of magnitude in the Palatini case. We summarize all relevant energy scales in table \ref{table2}.

	\begin{table}
	\begin{center}
		\begin{tabular}{c|ccccccc}
		& 	$\chi_{\text{inf}}$  & 	$\chi_{\text{jump}}$& 	$\chi_{\text{end}}$& 	$h_{\text{inf}}$& $h_{\text{jump}}$&	$h_{\text{end}}$&	$\mu_{\text{inf}}$ \\[8pt]
	\hline 	\\[-5pt]
		 Metric & $M_P\ln N$  & 	$\dfrac{M_P}{\xi}$  & $M_P$& $\dfrac{M_P}{\sqrt{\xi}} \sqrt{N}$ & 	$\dfrac{M_P}{\xi}$  &$\dfrac{M_P}{\sqrt{\xi}}$&  $y_t \dfrac{M_P}{\sqrt{\xi}}$ \\	\\
		 	Palatini 	& $\dfrac{M_P}{\sqrt{\xi}}\ln\xi N$  &  $\dfrac{M_P}{\sqrt{\xi}}$&$\dfrac{M_P}{\sqrt{\xi}}\ln\xi$& $M_P \sqrt{N}$&$\dfrac{M_P}{\sqrt{\xi}}$& $M_P$&  $y_t \dfrac{M_P}{\sqrt{\xi}}$
		\end{tabular}
		\caption{Comparison between the energy scales in metric and Palatini Higgs inflation.}
		\label{table2}
		\end{center}
	\end{table}

So far, the analysis has given us no information about how big or small the parameter $\delta\lambda$ can be. 
In order to estimate its order of magnitude, we study the 
 $\beta$-function of $\lambda$ in the presence of new physics. As an illustrative example, we first consider a new degree of freedom with mass $m\sim \Lambda$, but we assume that it is a spectator field, \ie it has no impact on inflation beyond changing the running of $\lambda$.  We also ignore all quadratic divergences associated with these particles (see \cite{Bardeen:1995kv, 0708.3550, 1303.7244}).
In the metric case, the resulting correction to the $\beta$-function scales as
\begin{equation} \label{deltaBetaMetric}
\delta \beta_{\text{metric}} = \frac{g^2}{16 \pi ^2} \,,
\end{equation}
where we assumed that the Higgs field couples to the new degree of freedom $\phi$ via the term $g h^2 \phi^2$.
Integrating this quantity from $m$ to $\mu_{\text{inf}}\sim y_t M_P/\sqrt{\xi}$, we get the deviation of $\lambda$ from the Standard Model prediction
\begin{equation} \label{deltaLambdaMetric}
	\delta \lambda_{\text{metric}} = \frac{g^2 \ln\xi}{32 \pi^2} \approx 2 \cdot 10^{-2} \,, 
\end{equation}
where we conservatively set $g=1$ and $\xi = 10^3$ for the estimate of the numerical value.
In contrast, the correction to the $\beta$-function in the Palatini case depends on the renormalization scale $\mu$ and is of the form\footnote
{One can obtain this result by expressing the one-loop contribution to the vertex function in terms of the physical coupling, which is defined in the momentum dependent subtraction scheme at the scale $\mu$ and thus accounts for decoupling of heavy degrees of freedom for $\mu \ll m$.} 
\begin{equation} \label{deltaBetaPalatini}
\delta \beta_{\text{Palatini}}= \frac{g^2}{16 \pi ^2} \frac{\mu^2}{m^2}\,.
\end{equation}
We used that, unlike in the metric theory, here \text{$\mu < m$}. Integrating this from $0$ to $\mu_{\text{inf}}\sim y_t M_P/\sqrt{\xi}$ yields
\begin{equation} \label{deltaLambdaPalatini}
\delta \lambda_{\text{Palatini}} = \frac{g^2 y_t^2}{32 \pi^2} \approx 6 \cdot 10^{-4}\,,
\end{equation}
where we again conservatively set $g\approx 1$.\footnote
{More precisely, the mass of new particles scales as $m \approx g M_p/\sqrt{\xi}$ for small $g$. This leads to $\delta \lambda_{\text{Palatini}} = y_t^2/(32 \pi^2)$ but does not change our numerical estimate.}
We note that since $y_t$ is of the order of $1$ (even at inflationary scales), the smallness of the numerical value in \Eq \eqref{deltaLambdaPalatini} is mainly caused by the loop suppression.\footnote
{Of course, $\delta \lambda_{\text{metric}}$ also contains a loop factor. But the difference is that in the metric case, the suppression is partly removed by the integration over a large energy range, which results in the factor $\ln \xi$.}
 Comparison of \Eqs \eqref{deltaLambdaMetric} and \eqref{deltaLambdaPalatini} provides an argument that Palatini Higgs inflation is considerably less sensitive to the contribution of new physics above $\Lambda$.

The goal of the above estimate is to highlight differences between metric and Palatini Higgs inflation within the same setup and assumptions. We stress that numerical values of $\delta\lambda$ can only be regarded as order-of-magnitude estimates, since prefactors depend on the precise nature of new physics at $\Lambda$. For example, this prefactor could become big if a large number of species existed at $\Lambda$. However, barring such scenarios we believe that our estimates, especially the value \eqref{deltaLambdaPalatini} in the Palatini scenario, represent conservative upper bounds. In particular, it is easy to imagine a situation in which the actual deviation of $\lambda$ is much smaller. This can be achieved if new physics does not directly couple to the Higgs and, consequently, only enters the running of $\lambda$ as a 2-loop effect. 

Coming back to UV-completion of the different Higgs inflation scenarios, we do not know what happens above $\Lambda$, \ie if new degrees of freedom appear or if the old fields merely enter the regime of strong coupling.
As explained above, however, the second option appears to be more likely within our setup of a minimal UV-completion, in which we assume that inflationary predictions are not changed at tree level. Therefore, it is important to investigate the robustness of the $\beta$-function in this case. To this end, we need to study the effect of
 higher-dimensional operators at the scale $\Lambda$ that respect the approximate shift symmetry $\chi \rightarrow \chi + $const of the potential $U(\chi)$. It turns out that they also lead to a correction of the form \eqref{deltaBetaPalatini} \cite{1308.2627, 1402.1476}. Consequently, we expect that the bound \eqref{deltaLambdaPalatini} is applicable in our setup.

\section{Connection of low- and high-energy physics}
\label{sec:connection}

So far, we have studied under which conditions the RG evolution computed within the Standard Model is robust against corrections due to new physics above $\Lambda$. 
In the present section, we assume that such corrections are small. This allows us to connect low-energy and inflationary parameters by extending the Standard Model running of the Higgs self-coupling $\lambda$ up to inflationary scales. In particular, we discuss bounds on $\xi$ following from measurements of the top Yukawa coupling $y_t^{\text{low}}$ at the weak scale, as well as bounds on $y_t^{\text{low}}$ due to inflation.
Finally, we study the self-consistency of our results by comparing $\lambda(\mu_{\text{inf}})$ to $\delta \lambda$ as computed in the previous section. In the following, we mainly focus on the Palatini case and only briefly address the metric scenario in the end.

Plugging the slow-roll parameter $\epsilon$ as defined in \Eq \eqref{slowRollParameters} in the power spectrum normalization \eqref{normalization} and using \Eq \eqref{NofH}, we deduce
\begin{equation} \label{xiOfLambda}
\xi = \frac{2 \lambda (N+1)^2}{5.0\cdot 10^{-7}} = 1.1\cdot 10^{10} \lambda \,,
\end{equation}
where we defined $h_{\text{end}}$ by the condition $\eta = -1$ and we plugged in $N=50.9$ in the last step. In this formula, $\lambda$ needs to be evaluated at inflationary energies.
Using the RG equations of the Standard Model, we compute $\lambda$ as a function of the renormalization scale $\mu$.\footnote
{We thank Fedor Bezrukov for kindly providing us with a script to do so. The script computes the running of the Standard Model couplings  in the $\overline{\text{MS}}$-scheme. It uses two-loop matching of physical and $\overline{\text{MS}}$ parameters  \cite{1507.08833} and three-loop RG evolution of coupling constants \cite{1205.2892} (see also \cite{1205.2893,1205.6497,1307.3536}). As an input, we use $\alpha_s(M_Z)=0.1181$ and $m_h=125.1\,$GeV. The influence of $\xi$ on the running is negligible \cite{0904.1537}, and we do not consider it.}
The result is then fitted with the well-known function
\begin{equation} \label{runningCoupling}
\lambda(\mu) = \lambda_0 + b \ln^2\left(\frac{\mu}{q M_P}\right) \,.
\end{equation}
In this way, the parameters of the Standard Model yield $\lambda_0\ll 1$, $b\sim 10^{-5}$ and $q\lesssim 1$. The main source of uncertainty in the values of these coefficients comes from measurements of the top Yukawa coupling at low energies. This makes it natural to parametrize all quantities as functions of $y_t^{\text{low}}$, which appears as an initial condition for the RG running towards high energies. As is common (see \cite{Tanabashi:2018oca}), we define it in the $\overline{\text{MS}}$-scheme. We use the normalization point $\mu = 173.2\,\text{GeV}$. Now we plug \Eqs (\ref{mu}) and \eqref{runningCoupling} in formula \eqref{xiOfLambda} and obtain
\begin{equation} \label{xiEquation}
\xi = 1.1\cdot 10^{10}\left\{\lambda_0 + b \ln^2\left[\frac{y_t}{\sqrt{\xi}q }\tanh \left(\dfrac{\sqrt{\xi}\chi}{M_P}\right)\right]\right\} \,.
\end{equation}
Here $y_t$ is the value of the top Yukawa coupling at high energies, whereas the dependence of $\lambda_0$, $b$ and $q$ on $y_t^{\text{low}}$ is implicit. We neglect the running of $y_t$ at inflationary scales and instead use the value \eqref{Yt}. Note that we have assumed that the dependence of $\lambda$ on $\chi$ does not significantly modify the slow-roll parameter $\epsilon$, \ie that the tree-level version of \Eq \eqref{NofH} is still approximately valid. The numerical analysis will confirm that this assumption holds in the whole viable parameter space.

\Eq \eqref{xiEquation} allows us to fix $\xi$ as a function of $y_t^{\text{low}}$. Before displaying the resulting solutions, we need to investigate their consistency. A necessary condition for successful inflation is that the derivative of the potential is positive up to the scale of inflation:\footnote
{In fact, one should impose the slightly stricter criterion of avoiding eternal inflation because otherwise the classical treatment is not valid. Since it will turn out in the numerical analysis that $\diff U/\diff \chi$  never comes close to $0$, this distinction is inessential.}
\begin{equation}
\left.\dfrac{dU}{d\chi}\right\vert_{\chi<\chi_{\text{inf}}}>0 \,.
\end{equation}
Note that this does not exclude that the Higgs potential has a stationary point at a higher energy. The condition can be expressed in terms of the $\beta$-function of $\lambda$:
\begin{equation} \label{runningCondition}
\lambda(\mu) > -\frac{1}{4}\mu \frac{\partial \lambda(\mu)}{
	\partial \mu}  \,, ~~~ \mu<\mu_{\text{inf}} \,,
\end{equation}
where we used \Eqs \eqref{Potential}, \eqref{mu} and approximated $F(\chi)$ by one after taking the derivative. This leads to the minimal value $\lambda_{\text{min}}$ that is still compatible with inflation:
\begin{equation} \label{lambdaLowerBound}
\lambda_{\text{min}}=	\frac{b}{4} \ln \left(\dfrac{q^2\xi}{y_t^2}\right)\,.
\end{equation}
Plugging this in \Eq \eqref{xiOfLambda}, we obtain\footnote
{This bound only refers to inflation on the plateau. Other scenarios, such as hilltop inflation, can allow for smaller values of $\xi$; see \cite{1709.07853, 1802.09299}.}
\begin{equation} \label{xiMinAnalytic}
\xi_{\text{min}}^{\text{analytic}} = 8.7 \cdot 10^5\,.
\end{equation}
This corresponds to $y_{t}^{\text{low}} = 0.92367$ and $\lambda(\mu_{\text{inf}})=8.0 \cdot 10^{-5}$. 
However, these conclusions are only valid if \Eq \eqref{NofH} is not strongly modified due to the running of $\lambda$. Close to the minimal value of $\xi$, this is no longer the case, so that numerical analysis is required, in the course of which it turns out that \Eq \eqref{xiMinAnalytic} has to be mildly corrected (see appendix \ref{app} for details):
\begin{equation} \label{xiMin}
\xi_{\text{min}} = 1.0 \cdot 10^6 \,,
\end{equation}
which gives $y_{t}^{\text{low}}=0.92368$ and $\lambda(\mu_{\text{inf}})=1.0 \cdot 10^{-4}$.\footnote
{A lower bound on $\xi$ has already been discussed in a different setup in \cite{1709.07853}. There, no connection was made to low-energy data of the Standard Model, i.e., $\lambda_0$, $b$ and $q$ were treated as free parameters. Moreover, only positive values of $\lambda_0$ were considered. This led to $\xi_{\text{min}}\approx 6\cdot 10^6$ \cite{1709.07853}.}

Let us now turn to the upper limit on $\xi$. It arises from the lower limit on the low-energy value $y_{t}^{\text{low}}$ of the top Yukawa coupling as determined by collider experiments. As input, we can use measurements of the top quark mass, such as the CMS-result \cite{1509.04044} $m_{t}^{\text{MC}} = 172.44\pm 0.60 \, \text{GeV}$ and the ATLAS-result \cite{1810.01772}  $m_{t}^{\text{MC}} = 172.69\pm 0.66 \, \text{GeV}$ (see also the review of the top quark in \cite{Tanabashi:2018oca}). The cited values correspond to Monte Carlo masses. To extract from them the Yukawa coupling, one proceeds as follows \cite{1411.1923}. First, one relates them to the top pole mass, which introduces an uncertainty of about $1\,$GeV (see, \eg \cite{1412.3649, 1801.03944}).  Thus, we consider the conservative limit $m_{t}^{\text{pole}} > 170\,\text{GeV}$. Secondly, one needs to account for strong and electroweak corrections to obtain the top mass $m_t^{\overline{\text{MS}}}$, which is related to the Yukawa coupling as $m_t^{\overline{\text{MS}}}=y_t^{\text{low}}v/\sqrt{2}$. Here $v=246.22\,\text{GeV}$ is the expectation value of the Higgs field. We arrive at the lower bound $y_t^{\text{low}}>0.91860$.
This gives $\lambda(\mu_{\text{inf}}) = 6.3\cdot 10^{-3}$ and hence
\begin{equation} \label{upperXi}
\xi_{\text{max}} = 6.8 \cdot 10^{7} \,.
\end{equation}
We must emphasize, however, that our goal here is not to give the precise number, but to point out that the lower bound on the top Yukawa coupling due to collider experiments leads to a fairly strict upper bound on $\xi$.

\begin{figure}
	\centering 
	\begin{subfigure}{0.6\textwidth}
		\includegraphics[width=\textwidth]{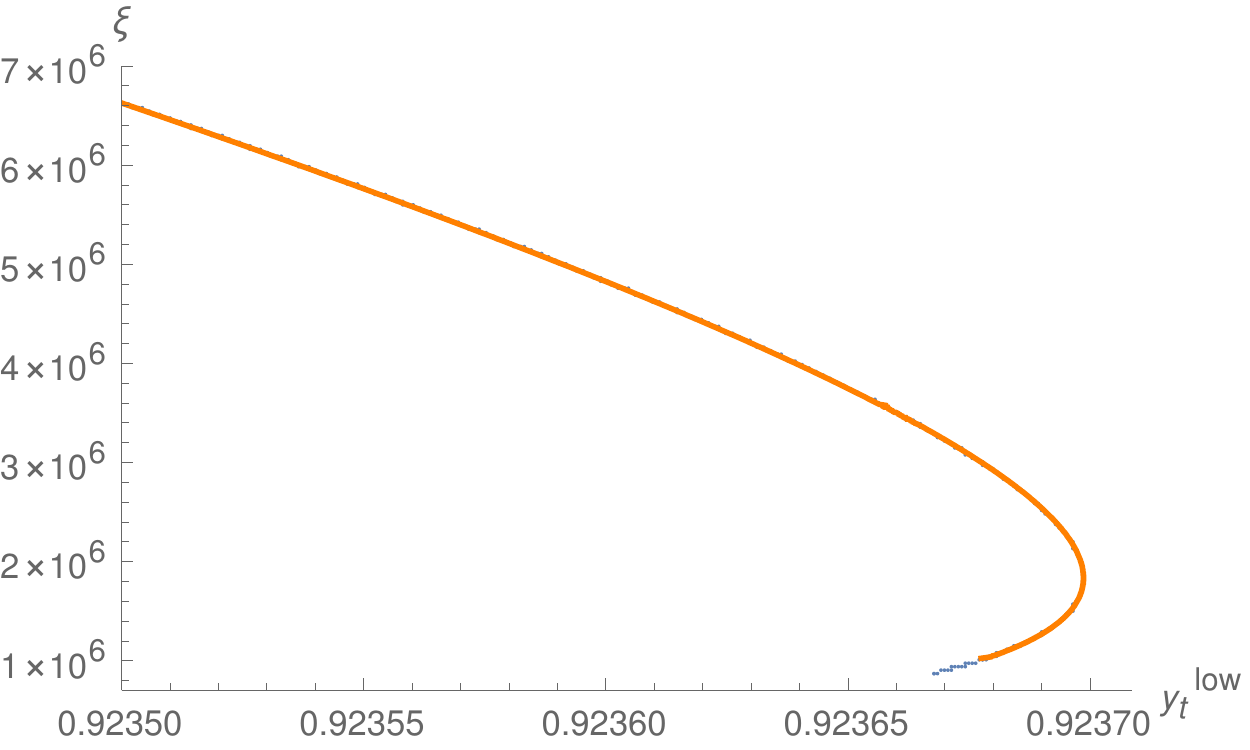}
	\end{subfigure}
	\hspace{0.02\textwidth}
	\caption{Non-minimal coupling $\xi$ as a function of the low-energy value of top Yukawa coupling $y_t^{\text{low}}$, following from analytic (in blue) and numerical (in orange) analysis. Close to the lower bound on $\xi$, the analytic estimate predicts the solution that is absent in the numerical calculation. Otherwise, the analytic and numerical results agree with each other. Smaller values of $y_t^{\text{low}}$ are also viable but not shown.}
			\label{sfig:xiOfLambda0}
\end{figure}

Now the following picture emerges for consistent solutions of \Eq \eqref{xiEquation}, which are displayed in \fig \ref{sfig:xiOfLambda0}. For sufficiently small values of $y_t^{\text{low}}$, there is a unique result for $\xi$. In an intermediate range of $y_t^{\text{low}}$, we observe that there are two distinct values of $\xi$ for the same $y_t^{\text{low}}$. For even bigger values of $y_t^{\text{low}}$, no solution exists any more. The biggest value for which we find a solution is 
\begin{equation} \label{topMassMax}
y_t^{\text{low, max}} = 0.92370 \,.
\end{equation}
It corresponds to $\xi = 1.8 \cdot 10^6$ and $\lambda(\mu_{\text{inf}}) = 1.7 \cdot 10^{-4}$. We remind the reader that $y_t^{\text{low}}$ is defined in the $\overline{\text{MS}}$-scheme at the renormalization point $\mu=173.2\,\text{GeV}$. In terms of the top pole mass discussed above, the bound \eqref{topMassMax} would correspond to $m_t^{\text{pole}}<170.9\,\text{GeV}$.

At this point, we must investigate the robustness of our findings against possible corrections from new physics at the cutoff scale $\Lambda$. Comparing the results for $\lambda(\mu_{\text{inf}})$ with our estimate \eqref{deltaLambdaPalatini}, we expect the upper bound \eqref{upperXi} to be robust. In contrast, both the lower bound \eqref{xiMinAnalytic} on $\xi$ and the bound on the top mass \eqref{topMassMax} appear to be sensitive to corrections. The reason is that the running of $\lambda$ as computed within the Standard Model drives it to zero at high energies and therefore even small correction can be significant.\footnote
{It is an interesting numerical coincidence that $\delta \lambda$, the magnitude of which is essentially given by a loop factor, and values of $\lambda$ that can support inflation are of the same order of magnitude.}
To estimate how strongly new effects at $\Lambda$ can alter the bound on the top mass, we redo the above analysis but replace $\lambda \rightarrow  \lambda \pm \delta \lambda$, where we use $\delta \lambda = 6\cdot 10^{-4}$ as given in \Eq \eqref{deltaLambdaPalatini}. This changes the bound \eqref{topMassMax} by about $0.0006$.  Therefore, we can conservatively conclude that 
\begin{equation} \label{topMassMaxResult}
y_t^{\text{low}}\lesssim 0.925\,.
\end{equation}
 In terms of the top pole mass, this would correspond to $m_{t}^{\text{pole}}<171.1\,\text{GeV}$.

Finally, we briefly study the dependence of our result on the slight ambiguity in the choice of the number of e-foldings $N$, the top Yukawa coupling $y_t$ as well as the value of $\eta$ used to define the end of inflation. Changing them by $\delta N = \pm 0.5$, $\delta y_t = \pm 0.02$ as well as the choices $\eta = -1/2$ and $\eta = -2$ alter our results \eqref{xiMin} and \eqref{topMassMax} by less than $0.1 \cdot 10^6$ and $0.00001$, respectively. Moreover, we remark that new physics at $\Lambda$ can also change the running of $y_t$, which leads to a jump $\delta y_t$. Its order of magnitude is also given by \Eq \eqref{deltaLambdaPalatini}, \ie $\delta y_t \ll 0.02$. Consequently, one can safely neglect this effect.

Next, we apply the same analysis to metric Higgs inflation i.e., we assume for a moment that there are no corrections due to new physics above $\Lambda$ and then search for the biggest value of $y_t^{\text{low}}$ that is compatible with inflation. This analysis has already been performed in \cite{0812.4950, 0904.1537} (before the discovery of the Higgs boson). 
Since in the metric scenario $\xi$ is smaller than in the Palatini case, the scale \eqref{mu} of inflation is higher. Therefore, metric Higgs inflation probes larger parts of the effective Higgs potential, \ie it is sensitive to the appearance of stationary points at higher scales. For this reason, we expect that the resulting bound will be sharper.

In order to make the argument quantitative, we use the data displayed in table \ref{table} to evaluate the potential $U$, the slow-roll parameter $\epsilon$ as well as \Eq \eqref{NofH}, and plug the results in the Planck normalization \eqref{normalization}. Then we get instead of \Eq \eqref{xiOfLambda}
\begin{equation}
\xi = \sqrt{\frac{\lambda}{3}}\frac{N+1}{\sqrt{5.0\cdot 10^{-7}}} = 4.6\cdot 10^4 \sqrt{\lambda} \,,
\end{equation}
where as in the Palatini case we defined $h_{\text{end}}$ by the condition $\eta=-1$ and we used $N=55.4$ in the second step.
Now we insert the running coupling \eqref{runningCoupling} with the energy scale $\mu_{\text{inf}}$ given in \Eq \eqref{mu}. For a fixed top Yukawa coupling $y_t^{\text{low}}$, this gives an equation that determines $\xi$. 

The resulting solutions are only valid if $\diff U/\diff \chi>0$, which leads to the same condition \eqref{runningCondition} (or, equivalently, \eqref{lambdaLowerBound}) as in the Palatini case.
For sufficiently small values of $y_t^{\text{low}}$, we find the known plateau solution with $\xi$ lying between $10^3$ and $10^4$. As $y_t^{\text{low}}$ increases, the value of $\xi$ decreases. The biggest value of $y_t^{\text{low}}$ to which the plateau solution can be continuously deformed is
\begin{equation}\label{topMassMaxMetric}
y_t^{\text{low, max}} = 0.92289 \,,
\end{equation}
which corresponds to $\xi = 300$ and $\lambda(\mu_{\text{inf}}) = 4.2\cdot 10^{-5}$.
As expected, the bound is stronger in the metric case. But since the Higgs potential is very sensitive to $y_t^{\text{low}}$, the difference to \Eq \eqref{topMassMax} only amounts to $0.0008$. We expect the uncertainty in the value \eqref{topMassMaxMetric} due to the choices of $N$, $y_t$ and $\eta$ to be as small as in the Palatini case. However, we have not performed a full numerical study of the system. Therefore, our conclusions are only valid if the solution that corresponds to \Eq \eqref{topMassMaxMetric} indeed exists beyond the analytic approximation. 

An important difference to the Palatini model becomes apparent when we study how strongly corrections above $\Lambda$ can influence the bound \eqref{topMassMaxMetric}. Comparing $\lambda(\mu_{\text{inf}})$ to our estimate of $\delta \lambda$ displayed in \Eq \eqref{deltaLambdaMetric}, we observe that $\lambda(\mu_{\text{inf}})\ll \delta \lambda$. In accordance with previous studies \cite{1008.5157, 1403.6078, 1412.3811}, we therefore conclude that, in general, low- and high-energy parameters are not connected in the metric Higgs inflation and that the bound \eqref{topMassMaxMetric} does not apply. In summary, the metric and the Palatini scenarios yield similar bounds on the top Yukawa coupling $y_t^{\text{low}}$. However, the crucial difference between the two models is that this bound is more robust against corrections due to new physics in the Palatini case.

\section{Discussion}
\label{sec:conclusion}
Higgs inflation exists in its original metric version \cite{0710.3755} and the Palatini variant \cite{0803.2664}. They lead to different predictions for inflationary observables, but so far observations of the CMB cannot distinguish between them. An argument that is put forward in favor of Palatini Higgs inflation is that its cutoff scale $\Lambda$, above which perturbation theory breaks down, is parametrically higher than in the metric scenario \cite{1012.2900}. However, also in the metric case all relevant energy scales (except for the preheating temperature) lie below the cutoff, once one takes into account that $\Lambda$ depends on the value of the background field \cite{1008.5157}. In the scenario of a minimal UV-completion, \ie when physics beyond the Standard Model does not modify inflation at tree level, both versions of Higgs inflation might therefore appear to be equally predictive. In other words, it has been unclear if the higher cutoff scale in the Palatini scenario bears any physical relevance.

In the present work, we have proposed that the increased value of $\Lambda$ allows one to connect inflationary observables to low-energy parameters measured in collider experiments. Unless new physics above $\Lambda$ leads to unnaturally large corrections, the Standard Model RG evolution can be extended to inflationary scales. The fact that inflation is only possible for values of the non-minimal coupling $\xi\gtrsim 10^6$ leads to an upper bound on the low-energy value  of the top Yukawa coupling: $y_t^{\text{low}} < 0.925$ (computed in the $\overline{\text{MS}}$-scheme at $\mu=173.2\,\text{GeV}$). In terms of the top pole mass, this corresponds to $m_{t}^{\text{pole}}<171.1\,\text{GeV}$. It is intriguing that recent measurements of the top quark are fully compatible with this bound \cite{1904.05237, 1905.02302}.\footnote
{CMS and ATLAS  have measured $m_{t}^{\text{pole}}=170.5\pm0.8\,\text{GeV}$ \cite{1904.05237} and $m_{t}^{\text{pole}}=171.1\pm1.2\,\text{GeV}$ \cite{1905.02302}, respectively.}
Conversely, collider experiments lead to restrictions on inflation. The lower bound on $y_t^{\text{low}}$ implies $\xi \lesssim 10^8$. This constrains the tensor-to-scalar ratio $r$ within two orders of magnitude.

One cannot exclude the possibility that corrections due to new physics at $\Lambda$ are strong enough to invalidate the connection of low- and high-energy parameters. Even more drastically, it could also turn out in the end that the UV-completion of the Palatini model leads to tree-level modifications. In such cases, the situation in Palatini Higgs inflation would be the same as in the metric scenario. Inflation becomes independent of collider measurements of $y_t^{\text{low}}$ and, in particular, values of $\xi$ below the bound (\ref{xiMin}) and above the bound (\ref{upperXi}) are no longer excluded.

Assuming that the connection of low- and high-energy physics can indeed be established, it is interesting to express the resulting bounds in terms of the value $\lambda_0$, which the running Higgs self-coupling assumes in its high-energy minimum. As is well-known, current measurements of $y_t^{\text{low}}$ favor negative values. In contrast, our analysis of inflation almost excludes all values below zero, $\lambda_0 \gtrsim -10^{-3}$. Thus, empirical data -- collider experiments and the amplitude of CMB fluctuations -- imply that $\lambda_0\approx 0$.  Before the measurement of the Higgs mass, this outcome has already been predicted from the ``multiple  point criticality principle'' \cite{Froggatt:1995rt} and in the context of asymptotic safety \cite{0912.0208}.

So far, we have not committed ourselves to a particular form of the UV-completion of the theory above $\Lambda$. In \cite{2001.09088}, we specialized to the case in which no new fundamental degrees of freedom exist anywhere above the weak scale. Moreover, we set the mass of the Higgs boson to zero. In this case, the action \eqref{action_full} (with $v=0$) does not only lead to successful inflation but can also address the hierarchy problem. First, there is no issue regarding the sensitivity of low-energy physics to new heavy particles, since the latter are absent. Secondly, it has been shown that a non-perturbative gravitational effect has the potential to generate electroweak symmetry breaking, thereby opening up the possibility to calculate the Higgs mass \cite{2001.09088}.

Other phenomena that cannot be explained within the Standard Model, such as dark matter, neutrino oscillations and baryon asymmetry of the Universe, can also be addressed without introducing new particles with masses above the weak scale. This is for example realized in the Neutrino Minimal Standard Model \cite{Asaka:2005an, Asaka:2005pn}, the particle content of which is enlarged compared to that of the Standard Model only by three Majorana neutrinos. Finally, also gravity may be self-complete without any need for additional degrees of freedom \cite{1005.3497, 1010.1415, 1103.5963}. Thus, it might turn out that a more complete description of Nature requires fewer ingredients than often assumed.

\section*{Acknowledgments}

The authors thank Fedor Bezrukov and Javier Rubio for useful comments on the draft of the manuscript. The work was supported by ERC-AdG-2015 grant 694896 and by the Swiss National Science Foundation Excellence grant 200020B\underline{ }182864.

\appendix

\section{Numerical lower bound on $\xi$}
\label{app}

The analytic analysis of section \ref{sec:connection} is no longer trustable as soon as the running of $\lambda$ leads to a sizable modification of $\diff U/\diff \chi$. The main complication comes from the fact that \Eq \eqref{NofH} changes in such a way that $h$ is no longer independent of $\xi$. Consequently, the equations that determine $h$ and $\xi$ become coupled and more difficult to solve.
For the numerical study, we tackle this problem in the following way. For a given $y_t^{\text{low}}$, we guess a value of $h$. Then the only unknown variable in \Eq \eqref{normalization} is $\xi$ and we can solve for it. Subsequently, it is straightforward to compute $h_{\text{end}}$ (corresponding to $\eta=-1$) and $N$. If the resulting number of e-foldings is not sufficiently close to $50.9$ (see table \ref{table}), we repeat this procedure for a different value of $h$. This calculation does not rely on any approximation.

\begin{figure}
	\centering 
	\begin{subfigure}{0.6\textwidth}
		\includegraphics[width=\textwidth]{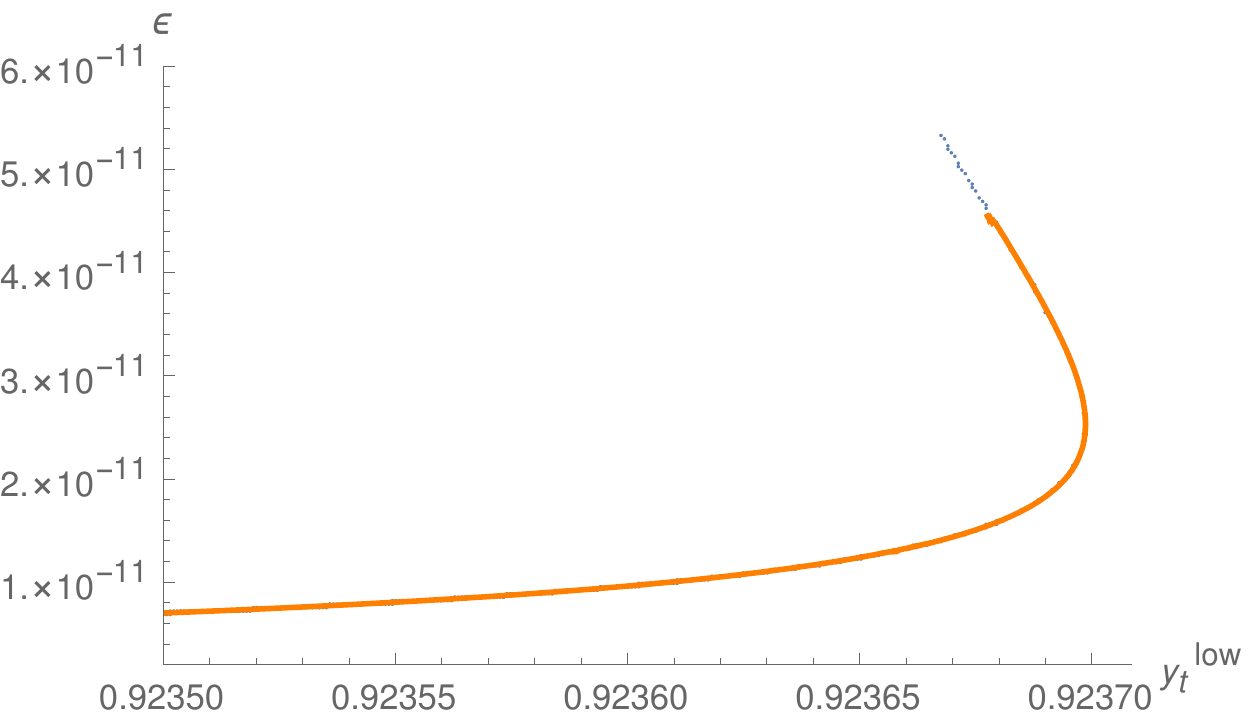}
	\end{subfigure}
		\caption{The slow-roll parameter $\epsilon$ as a function of the low-energy value of top Yukawa coupling $y_t^{\text{low}}$, following from analytic (in blue) and numerical (in orange) analysis. The upper part of the curve corresponds to the lower part of \fig \ref{sfig:xiOfLambda0}.}
				\label{sfig:epsilonOfLambda0}
\end{figure}

Using the above procedure, we determine $\xi$ for different values of $y_t^{\text{low}}$.\footnote
{We use Mathematica \cite{mathematica} for the numerical analysis.}
It turns out that in the whole parameter space, the full numerical result and the analytic approximation, which arises from solving \eqref{xiEquation} (in the approximation $\tanh \left(\sqrt{\xi}\chi/M_P\right)\approx 1$), agree perfectly. The only difference is that the numerical result stops existing for a higher value of $\xi$, as displayed in \Eq \eqref{xiMin}.
At this point, the Higgs potential does not yet develop a stationary point at inflationary energies, but it appears that it no longer allows for a sufficient number of e-foldings.  In \fig \ref{sfig:xiOfLambda0}, both the analytic and numerical results are shown, and we restricted ourselves to the most interesting region of large $y_t^{\text{low}}$. Most importantly, the numerical study confirms the upper bound \eqref{topMassMax} on the top Yukawa coupling.

Moreover, we compute the inflationary indices. \fig \ref{sfig:epsilonOfLambda0} shows the numerical result for $\epsilon$ as well as the analytic estimate $\epsilon = 1/(8 \xi (N+1)^2)$. Again, the two agree perfectly. This explains why \Eq \eqref{xiOfLambda} yields the correct result for $\xi$. 
Finally, $\eta$ (and consequently $n_s$) are independent of $y_t^{\text{low}}$, as expected (see table \ref{table}), and the analytic and numerical results agree.

\bibliographystyle{JHEP}
\bibliography{inflationBib}

\end{document}